\documentclass[sigconf,screen]{acmart}

\usepackage{balance}
\AtBeginDocument{%
  \providecommand\BibTeX{{%
    \normalfont B\kern-0.5em{\scshape i\kern-0.25em b}\kern-0.8em\TeX}}}

\begin{document}

\title{Customer Service Representative's Perception of the AI Assistant in an Organization's Call Center}

\author{Kai Qin}
\email{kyle.qin@hotmail.com}
\affiliation{%
  \institution{Guangxi Power Grid Co., Ltd.}
  \city{Nanning}
  \country{China}}

\author{Kexin Du}
\email{dkx22@mails.tsinghua.edu.cn}
\orcid{0009-0001-9666-8173}
\affiliation{
  \institution{Tsinghua University}
  \city{Beijing}
  \country{China}
}

\author{Yimeng Chen}
\affiliation{%
  \institution{Beijing Normal University}
  \city{Beijing}
  \country{China}}
\email{276082120@qq.com}
\orcid{0009-0006-9251-8641}

\author{Yueyan Liu}
\affiliation{%
  \institution{China Academy of Art}
  \city{Hangzhou}
  \country{China}}
\email{yueyanliu.hci@gmail.com}
\orcid{0009-0002-8591-1374}

\author{Jie Cai}
\authornote{Corresponding author}
\email{jie-cai@mail.tsinghua.edu.cn}
\orcid{0000-0002-0582-555X}
\author{Zhiqiang Nie}
\email{nzq24@mails.tsinghua.edu.cn}
\orcid{0000-0002-0582-555X}
\affiliation{%
  \institution{Tsinghua University}
  \city{Beijing}
  \country{China}}

\author{Nan Gao}
\affiliation{%
  \institution{Nankai University}
  \city{Tianjin}
  \country{China}}
\email{nan.gao@nankai.tsinghua.edu.cn}
\orcid{0000-0002-9694-2689}

\author{Guohui Wei}
\email{262598883@qq.com}
\author{Shengzhu Wang}
\email{215034471@qq.com}
\affiliation{%
  \institution{Guangxi Power Grid Co., Ltd.}
  \city{Nanning}
  \country{China}}
  
\author{Chun Yu}
\affiliation{%
  \institution{Tsinghua University}
  \city{Beijing}
  \country{China}}
\email{chunyu@tsinghua.edu.cn}
\orcid{https://orcid.org/0000-0003-2591-7993}

\renewcommand{\shortauthors}{Qin et al.}

\begin{abstract}
The integration of various AI tools creates a complex socio-technical environment where employee-customer interactions form the core of work practices. This study investigates how customer service representatives (CSRs) at the power grid service customer service call center perceive AI assistance in their interactions with customers. Through a field visit and semi-structured interviews with 13 CSRs, we found that AI can alleviate some traditional burdens during the call (e.g., typing and memorizing) but also introduces new burdens (e.g., earning, compliance, psychological burdens). This research contributes to a more nuanced understanding of AI integration in organizational settings and highlights the efforts and burdens undertaken by CSRs to adapt to the updated system.
\color{red}{Preprint Accepted at ACM CSCW Poster 2025}
\end{abstract}



\maketitle

\section{Introduction}

AI has increasingly reshaped human-computer interaction and is widely adopted in customer service organizations to automate repetitive tasks, improve efficiency, and support personalized customer interactions.
Recent studies have developed conceptual frameworks to understand how AI influences customer experiences throughout the customer journey, emphasizing its potential to augment human capabilities rather than replace them \cite{Integrating_Ying, inavolu2024exploring}. 
Some studies have revealed both the promise of AI to enhance task coordination and challenges that emerged during implementation, particularly around issues of trust, transparency, and cross-functional alignment, indicating a gap between what AI technology could do and how it was actually implemented in complex organizational settings \cite{ItMightbe, lim2022adoption, AutomationAdoptioninOrganizationsRemainaFallacy}.
There is a need to balance technological efficiency with human interaction to maintain effective customer relationships, particularly in service contexts with emotional labor \cite{libai2020brave}. However, they tend to prioritize customer experience or organizational benefits with strategic or managerial perspectives, while paying comparatively less attention to how frontline employees experience, interpret, and adapt to integrating AI assistants in their daily work.

The service sector presents unique challenges for AI integration due to its emphasis on direct customer engagement and emotional labor. Socio-technical factors, such as employee resistance and organizational culture, significantly shape how workers engage with or reject AI technologies \cite{UnderstandingSocio_technicalFactors, MakingofPerformativeAccuracy}. Furthermore, specialized professional roles present unique challenges for technological integration that must be addressed through careful socio-technical design \cite{CollaborativeWorkinMalwareAnalysis}. 
Workplace stress complicates AI implementation, as high-pressure environments can exacerbate employee concerns about job security and technological complexity \cite{UnderPressure}. 

Customer service representatives (CSRs) in service organizations face particularly complex challenges. On one hand, they are in a specialized role that requires specific skills and training to effectively use AI assistants. On the other hand, their work heavily relies on emotional engagement and complex interpersonal interactions, placing higher demands on the adaptability and human-centered design of the technology. At the same time, customer service is inherently high-pressure, with demands such as high call volumes, strict response time limits, and expectations for quick issue resolution, all of which can affect employees' willingness and ability to adopt new technologies. 

Power grid customer service call center offers a compelling context for studying AI integration due to their critical role in managing infrastructure services that directly impact citizens' daily lives \cite{huang2021perspective, ManagementCooperationSustainability}. CSRs require technical expertise and interpersonal skills to address customer inquiries while adhering to management directives and cooperative work processes. Integrating AI tools, such as large language models (LLMs), emotion recognition, and speech-to-text technologies, creates a complex socio-technical environment where employee perceptions significantly influence adoption and effectiveness. The involvement and perceptions of end-users, such as CSRs, in the testing and auditing AI systems have been underscored as critical for ensuring accountability and ameliorating potential burdens towards end-users \cite{deng2023supporting}. Therefore, our research focuses on CSRs in the power grid customer service call center, aiming to explore the following research question: \textbf{How do CSRs perceive AI assistance in their interaction with customers?}

\section{Method}

\begin{figure*}[tph]
    \centering
    \includegraphics[width=0.7\linewidth]{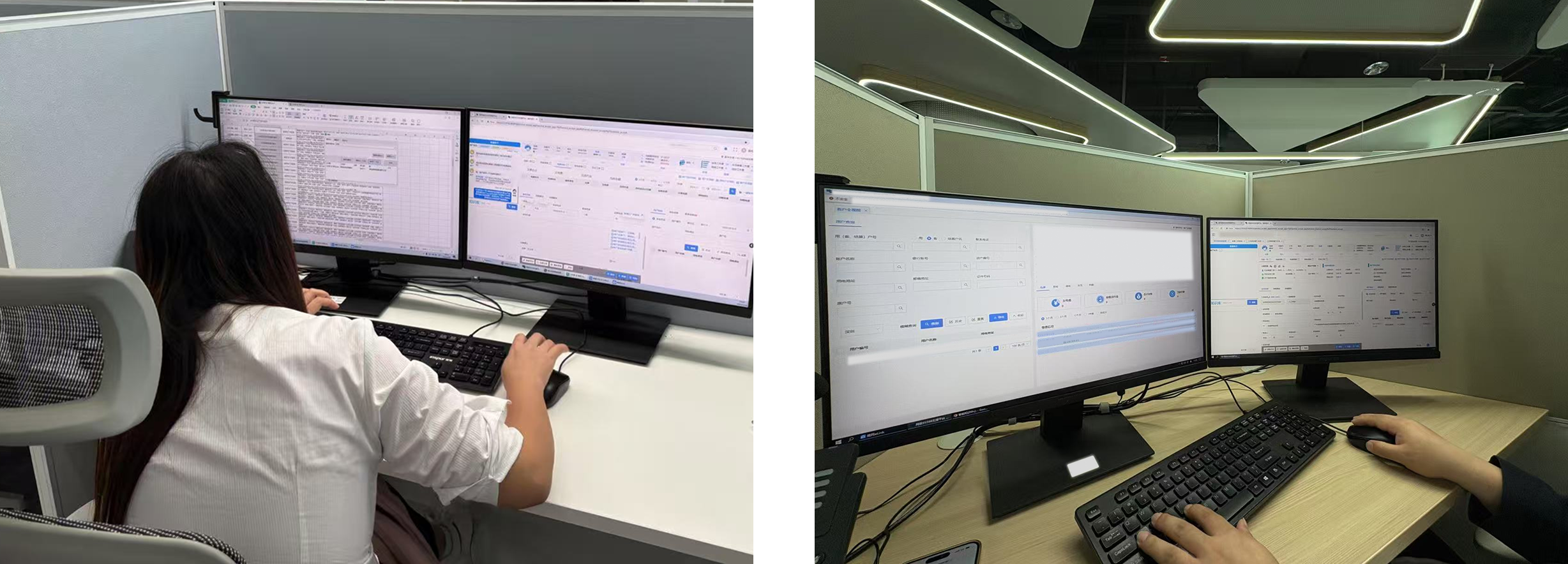}
    \Description{Each customer service representative (CSR) is provided with dual-monitor setups. One monitor is dedicated to the telephony system, enabling call handling and work order forms documentation, while the other is used to access commonly utilized tools, such as the internal knowledge management system and customer information databases. CSRs frequently switch among multiple software applications throughout the service process to gather necessary information and effectively address customer inquiries.}
    \caption{Typical Work Environment of Customer Service Representatives}
    \label{fig:work_environment}
\end{figure*}

\subsection{Interview and Data Analysis}

We visited the organization for a week to observe their workstations and discuss their workflow with department heads and directors. The observation helped us understand their organizational structure. We then conducted semi-structured interviews with 13 customer service representatives to gather their perceptions of the newly introduced AI assistants. \autoref{fig:work_environment} shows a typical work environment of CSRs. Each CSR has a dual-monitor setup (one for the telephony system, enabling call handling and form documentation, another for commonly utilized tools, such as the knowledge and customer management databases). CSRs frequently switch among multiple software applications throughout the service process to gather necessary information and effectively address customer inquiries. 
A typical workflow is as follows: CSRs answer calls through the system, ask the customer for basic information, look up relevant information and standardized responses based on the customer's issue, provide standardized replies to help resolve the problem, and fill out a form (e.g., name, meter ID, phone number, address, reason to call etc.) in the database to archive the issue. 

We conducted 13 semi-structured interviews (demographic details in \autoref{A:demo}) with CSRs in their work environment, incorporating observations of their computer screens, and asked them to demonstrate specific tasks when necessary. All 13 participants were hotline CSRs responsible for answering calls from users in specific regions. CSRs are categorized into roles based on their responsibilities: regular CSRs, team leaders, and shift supervisors. Among those we interviewed, P2 and P9 were team leaders, while the remaining participants were all regular CSRs.

The interview protocol included general questions about their workflow and the system interface, interactions with customers using the AI assistant, and demographic and work experience questions. We utilized a free open-source tool for collaborative coding, following the thematic analysis steps \cite{clarke2017thematic}. Research positionality is in \autoref{A:p}. Two researchers began by independently reviewing all transcripts and coding one selected transcript. They held weekly meetings to discuss and define codes and developed an initial codebook. One researcher subsequently coded the remaining transcripts, with a second reviewing for consensus. Since intercoder reliability is not commonly used in interviews \cite{10.1145/3359174}, researchers discussed the coding results through weekly meetings until all issues of disagreement were solved. After coding, the researchers organized the codes into themes and subthemes on a whiteboard during a group meeting. Finally, they compiled quotes in a shared document, categorizing them according to the research question. Our findings are mainly from the interviews.

\section{Findings}
\subsection{AI Detection Worked But With Challenges of Inaccuracy} 
AI could identify audio effectively and translate it to text in real time in interaction. CSRs appreciated the original version of the AI assistant design. However, they also reported inaccurate and unnecessary features, which could be annoying during the interaction. For example,  P1 stated, ``\textit{The response is almost instantaneous, with just a one-second delay.}'' Similarly, P3 noted that the AI's fast detection speed helped her understand the context during long conversations, but also mentioned issues with its performance on accents and technical cutoffs:
``\textit{Yes, it does help; sometimes customers speak very quickly, and it's hard to catch what they're saying, but the system can still recognize their speech. However, if the customer has a strong accent or speaks Cantonese, it often fails to recognize the input. Another issue is that sometimes parts of the audio are missing when the call comes in, so they can't be recognized at all. Still, it's useful—especially for longer calls, you can scroll through and get a general sense of what the customer was talking about.}''

CSRs actively employed AI-recognized information to enhance operational accuracy and efficiency, particularly when customers delivered information rapidly. P4 acknowledged that he relied on AI-generated transcripts to remind him of the detailed information during the communication process, which helped him offload the typing and memorizing burden:
``\textit{When a customer comes in and tells me they're from a certain residential area in XXX, I can't always remember the full address right away. So I scroll up to see the address they mentioned earlier. I rely on those cues to help me select the correct address and locate the specific meter. This feature is very useful. I really can't do without it. Sometimes, when customers report their ID numbers really quickly, there are also some prompts available to help. So it basically acts as some kind of note-taking assistant. I mean, I sometimes open a separate document to manually jot down customer information, but this note-taking assistant is still the main one I rely on, especially when it comes to things like phone numbers, addresses, or other important details. If I'm slow to record them, this note-taking assistant becomes very helpful.}''


However, the accuracy of transcripts relied heavily on clear pronunciation. The AI assistant has significant deficiencies in processing and displaying complete numerical sequences, especially with phone numbers. P1 reported that AI systems provided phone numbers in an inconsistent way and stated that it can only generate numbers one by one: ``\textit{AI assistant isn't that smart in reality. It gives phone numbers in bits and pieces, so I have to manually enter them.}'' Additionally, P5 pointed out that ``\textit{it may struggle; sometimes it confuses homophones, or only transcribes part of what was said.}''

The AI assistant's speech recognition was restricted by processing lengthy and complex conversations, leading to fragmented transcripts. P3 mentioned the cutoff issue during prolonged calls:
``\textit{I feel like it could record more, because with longer voice calls, it seems to stop capturing after a certain point. For example, when the conversation gets more intense, like if the customer has a lot of follow-up questions or is emotionally charged about an issue, the call might go on for over 30 minutes. But the system might only transcribe the first 10 or 15 minutes. After that, it just freezes or stops recording, and you don't get a complete transcription of the entire call. }''

The AI assistant offered emotion recognition features. However, CSRs revealed shortages of AI emotion recognition, including misclassifying normal speech intensity as negative emotions and lacking sufficient tags in emotion categories. Due to reliability concerns, CSRs largely disregarded emotion analysis features. P3 pointed out AI often gave false alerts about customers' attitudes: ``\textit{While it prompts CSRs about customer tone issues, it merely judges by volume level, often misinterpreting normal volume as attitude problems.}'' Thus, most CSRs would ignore the tags and state that they could easily recognize the customer's emotion during the call.

\subsection{AI-Assisted Form Completion Enhanced Productivity but Introduced Redundancy}

While reducing basic typing labor, AI-generated outputs introduced structural inefficiencies in information processing because most AI-prefilled content required manual correction or deletion. The AI prefilling function helped CSRs organize call records to a certain extent. However, the redundant content also required additional labor.

CSRs commented on the dynamic yet limited usefulness of the AI assistant's template generation feature during calls. As described by P2, ``\textit{During the call, the assistant keeps updating and generating content. Sometimes in the middle of the call or after it ends, it might revise or update the draft based on the full conversation.}'' While this functionality offered a real-time drafting aid, its practical utility remained constrained by its generic format as P2 added: ``\textit{But mostly, what it gives us is a preset template—it's not very detailed to the point that you can use it directly. We still have to do a lot of editing. It basically just gives you a rough template based on the customer's needs.}''  

In many cases, the AI-generated text required extensive trimming and rewording. As P2 further explained, ``\textit{Of course, if a brief description is enough to clearly express the customer's issue, we prefer a short version. Sometimes, we can describe the situation in just one or two sentences. Using the template takes more time—we have to trim it, cut unnecessary parts, and by the time I finish editing, I might've done it faster by just typing it myself.}'' Additionally, the AI assistant could prefill multiple forms simultaneously, but it may create visual clutter when handling complex service requests. These issues highlighted a common trade-off in AI-human collaboration: while automation assists in generating content, it can also impose additional cognitive and editing workload when its output fails to align with real-world task constraints.

AI transcription was convenient, but it could not extract key information. P3 described the difficulty of looking for certain information throughout the long conversation history: ``\textit{The AI assistant shows limitations during long or dialect-heavy calls. For instance, when a customer calls to check more than 20 electricity meters, the system only shows the current query while clearing previous records. I have to manually scroll up to review prior queries. It would greatly improve efficiency if the system could retain temporary records of current-call search history, even just basic caching.}''

CSRs reported that while the AI assistant's prefill feature supported form completion, it often failed to meet the accuracy and flexibility required for their tasks, especially in handling the address box and the power supply unit identification box. P2 explained: ``\textit{No matter what, once I click on that field, a prefill window will pop up, and we have to go through several steps to select the address, then add detailed address information manually. The system also auto-generates a corresponding power supply unit based on the address, which directly affects the accuracy of the later task assignment.}''

Despite the convenience, CSRs emphasized the necessity of manual verification due to frequent mismatches between the AI's output and organizational standards: ``\textit{If we don't manually verify and adjust it, and just go with the system's default suggestion, there will be many errors,}'' P2 added. Furthermore, the design of the system interface introduces operational friction—even when corrections are needed, ``\textit{it still brings up that same interface,}'' and they ``\textit{ have to make changes through that specific window}'', which slows down their workflow and reduces efficiency.

The AI assistant's transcription capability became especially critical when audio devices malfunctioned, as it helped prevent service interruptions. P4 shared a critical incident: ``\textit{Once when my headset malfunctioned, I could speak but couldn't hear the customer. The assistant's real-time transcription became a lifeline—I comprehended requests by reading on-screen dialogue and responded normally. In such emergencies, the assistant effectively served as a hearing substitute, preventing awkward customer delays.}''
This case revealed AI's capacity in emergency management by helping CSRs handle equipment failures and ensure customer satisfaction.

\section{Discussion}

Our findings reveal that an AI assistant can alleviate some traditional burdens, such as cumbersome and time-consuming tasks, thereby demonstrating potential for enhancing foundational efficiency, aligning with previous research in areas such as entertainment, work, education, and healthcare \cite{wang2019human, li2024flowgpt, purohit2023chatgpt}. However, it also introduces new learning requirements, compliance challenges, and psychological burdens. We identified the efforts that CSRs must invest in to adapt to AI assistants. Future research should address adaptation challenges, especially as these operational burdens and user experiences may influence their trust in AI assistants \cite{hauptman2022components,goyal2024designing}.
Specifically, we identified three types of burdens introduced by AI intervention in the organization's call center context \cite{aljuneidi2024fine, moynihan2015administrative, herd2019administrative}.  


Learning burden refers to the extra cognitive and operational load CSRs face from using the AI assistant. These include the AI's inability to accurately recognize phone numbers as a continuous string of digits, incorrect address recognition, or poor handling of dialects \cite{wasi2024diaframe}. These factors require CSRs to spend additional effort understanding or adapting to the system, thereby contributing to the learning burdens. On one hand, introducing the AI system has improved the efficiency of information recognition, especially when customers babble; the AI can accurately capture relevant details. However, when dialects or unclear pronunciations are involved, the system fails to properly retain or interpret the information.
On the other hand, the system exhibits deficiencies in telephone number processing, the integrity of long conversation transcription, and the retrieval of specific information from extensive dialogue histories. Thus, while the AI enhances work efficiency, it simultaneously increases CSRs' learning burdens due to the need for extra adaptation and correction. The mismatch between technological expectations and actual implementation reflects a common oversight among technology designers, who overestimate efficiency gains while underestimating the implicit learning burdens of adapting to new systems \cite{gomez2025does}. This phenomenon exists in the workplace and manifests in personal AI use in different forms \cite{namvarpour2024uncovering}.


Compliance burden refers to the effort CSRs require to align with institutional procedures, regulatory standards, and the organization's internal lexicon when completing tasks such as filling out work order forms. When CSRs use AI assistants, in addition to direct functional problems, they also implicitly bear a compliance burden. For example, when the prefilled content generated by AI is inaccurate or redundant, employees must manually correct, delete, and edit the content to ensure accuracy and standardization. Moreover, due to AI recognition's limited accuracy, employees must conduct manual verification to reduce the risk of errors, which constitutes a compliance burden for meeting quality requirements. These additional operations not only increase the workload but also reflect the implicit burdens incurred when employees adapt to and ``accommodate'' the imperfections of the AI system \cite{boubin2017quantifying}. The introduction of the AI assistant has, in many cases, increased this burden, as the content generated by LLMs tends to lack compliance with established procedural norms. For instance, when summarizing a customer's demand for inquiry, the AI generates verbose, natural language descriptions.
In contrast, CSRs are expected to use domain-specific terminology that reflects standardized categories embedded in the organization's workflow -- terminology that has been institutionalized through long-term operational practice and internal documentation. As a result, although the AI-generated content may provide a helpful starting point, agents are frequently required to reinterpret, restructure, and rephrase the output to comply with formal guidelines \cite{kim2024conditions}. This process reflects a tension between the flexibility of generative systems and the rigid compliance requirements of organizational information systems, ultimately contributing to an increase in compliance-related workload and institutional friction.


Psychological burden refers to the emotional and attitudinal experiences of CSRs when using the AI assistant during interactions with users. Overall, participants generally expressed positive attitudes toward the AI assistant, as it effectively reduced their cognitive load, particularly in remembering addresses and customer information \cite{schmidhuber2021cognitive}. However, the AI assistant also generated redundant or inaccurate content at times. For instance, when it produced multiple versions of the reply or failed to assist efficiently in form filling, participants reported that the extra information became a burden, interfering with their workflow by obscuring essential content. Under high workload or stress, these interruptions were perceived as frustrating, contributing to psychological burden. Whether the overall psychological burden is reduced or increased remains unclear and warrants further empirical investigation, particularly under varying task loads and user stress conditions.





\section{Limitations and Future Work}

Our study has several limitations, particularly concerning sample breadth. We collected data from 13 CSRs, all from a single regional centre, which means the insights are confined to the specific organisation model, workflow norms, and customer demographics of that centre, and may not generalise to other customer-service centres \cite{grondal2025organizational, batt2009globalization}. The single-site scope also raises the possibility of Hawthorne effects \cite{schwartz2013hawthorne}. Future studies on CSR-related work could broaden the sample context by adding additional sites or explicitly comparing role variations among CSRs. Another limitation relates to the evaluation of the Emotion-AI assistant. Our study reveals that CSRs largely ignore the assistant’s emotion classifier; however, we did not include log-based workflow data in our analysis. This omission may obscure the actual patterns of AI engagement and the reasons emotion-AI features are overlooked \cite{dumais2014understanding}. Future work could consider using quantitative log analysis (e.g., comparing alert frequency and CSR override rates) to substantiate these claims.



\balance
\bibliographystyle{ACM-Reference-Format}
\bibliography{main}

\appendix
\section{Researcher Positionality}
\label{A:p}
Our team included members from the customer service department and academic researchers. The leading author is like a scientist in the department who is very familiar with the customer service processes and can help scholars better understand the work steps. Scholars who are second-authors, third-authors, etc., had no background in customer service work experience. The scientist's involvement may bias findings to a certain extent, limiting our research scope. Thus, researchers outside the customer service team performed the data collection and analysis independently, which helped avoid the mentioned bias.

\section{Demographic Table}
\label{A:demo}
\begin{table}[h]
\caption{Demographics of Participants}
  \label{tab:demo}
    \begin{tabular}{cccc}
    \toprule
    PID  & Age & Gender & Experience\\
    \midrule
    P1  & 28  & M  & 4 months\\
    P2  & 33  & M   & 10 years \\
    P3  & 24  & M   & 1 year \\
    P4  & 21  & F   & 4 months \\
    P5  & 35  & M   & 8 years \\
    P6  & 25  & M   & 3 years \\
    P7  & 21  & M   & 3 months \\
    P8  & 23  & F   & 4 months  \\
    P9  & 30  & M   & 4 years \\
    P10 & 23  & M   & 1 year \\
    P11 & 30  & F   & 1 year \\
    P12 & 26  & F  & 3 months \\
    P13 & 25  & M   & 3 months \\
    \bottomrule
  \end{tabular}
\end{table}

\end{document}